\documentclass[aps,prb,twocolumn,superscriptaddress,floatfix]{revtex4-1}
 
\usepackage{graphicx}% Include figure files
\usepackage{dcolumn}% Align table columns on decimal point
\usepackage{verbatim}
\usepackage[colorlinks=true]{hyperref}% add hypertext capabilities
\usepackage{multirow}
\usepackage{bm}% bold math
\usepackage{times}
\usepackage{color}
\usepackage{amstext}
\usepackage{amsmath}
\usepackage{amssymb} 
\usepackage{amsfonts}
\usepackage{natbib}
\usepackage[final]{microtype}
\usepackage{csquotes,lmodern}
\usepackage{hyperref}
\hypersetup{hidelinks}
\newcommand{\ci}[1]{$_{\text{#1}}$} %Lower index in chemical formulae - e.g. Bi\ci{2}Se\ci{3} gives formulae for Bi2Se3
\newcommand{\de}{${}^{\circ{}}${}}

\newcommand{\TNSe}{Ta$_2$NiSe$_5$}  
\newcommand{\TNS}{Ta$_2$NiS$_5$}

\begin{document} 

\title{Infrared phonon spectra of quasi-one-dimensional \texorpdfstring{Ta$_2$NiSe$_5$}{Ta2NiSe5} and \texorpdfstring{Ta$_2$NiS$_5$}{Ta2NiS5}}

\author{T.~I.~Larkin}
\email{T.Larkin@fkf.mpg.de}
\affiliation{
Max Planck Institute for Solid State Research, Heisenbergstra{\ss}e~1,
70569 Stuttgart, Germany}

\author{R.~D.~Dawson}
\affiliation{
Max Planck Institute for Solid State Research, Heisenbergstra{\ss}e~1,
70569 Stuttgart, Germany}

\author{M.~H\"oppner}
\affiliation{
Max Planck Institute for Solid State Research, Heisenbergstra{\ss}e~1,
70569 Stuttgart, Germany}

\author{T.~Takayama}
\affiliation{
Max Planck Institute for Solid State Research, Heisenbergstra{\ss}e~1,
70569 Stuttgart, Germany}
\affiliation{
Institute for Functional Matter and Quantum Technology, University
of Stuttgart, Pfaffenwaldring 57, 70550 Stuttgart, Germany}

\author{M.~Isobe}
\affiliation{
Max Planck Institute for Solid State Research, Heisenbergstra{\ss}e~1,
70569 Stuttgart, Germany}

\author{Y.-L.~Mathis}
\affiliation{
Institute for Beam Physics and Technology, Karlsruhe Institute of Technology, 76344
Eggenstein - Leopoldshafen, Germany}

\author{H.~Takagi}
\affiliation{
Max Planck Institute for Solid State Research, Heisenbergstra{\ss}e~1,
70569 Stuttgart, Germany}
\affiliation{
Institute for Functional Matter and Quantum Technology, University
of Stuttgart, Pfaffenwaldring 57, 70550 Stuttgart, Germany}
\affiliation{
Department of Physics, The University of Tokyo, Hongo, Tokyo 113-0033,
Japan}

\author{B.~Keimer}
\affiliation{
Max Planck Institute for Solid State Research, Heisenbergstra{\ss}e~1,
70569 Stuttgart, Germany}

\author{A.~V.~Boris} 
\email{A.Boris@fkf.mpg.de}
\affiliation{
Max Planck Institute for Solid State Research, Heisenbergstra{\ss}e~1,
70569 Stuttgart, Germany}

%\date{\today}

        \begin{abstract}
                Using a combination of infrared ellipsometry, time-domain terahertz spectroscopy, and far-infrared reflectometry we have obtained the $ac$-plane complex dielectric function of monoclinic ($C2/c$) \TNSe{} and orthorhombic ($Cmcm$) \TNS{} single crystals.  The identified dipole-active phonon modes polarized along $a$ and $c$ axes  are in good agreement with density functional theory calculations. With increasing temperature the $a$-axis phonon modes of \TNSe{} become poorly discernible, as they are superimposed on the electronic background which gradually fills the energy gap near the monoclinic-to-orthorhombic phase transition temperature $T_c$ = 326 K. In \TNS{}, which does not exhibit such a structural transition and remains orthorhombic down to low temperatures, the $a$-axis phonon modes are superimposed on a persistent broad electronic mode centered near 16~meV. We attribute this difference to strongly overlapping exciton-phonon complexes in \TNSe{}, as opposed to isolated instances of the same in \TNS{}, and find this to be in good agreement with an excitonic insulator state below $T_c$ in the former, as compared to the absence of one in the latter.
        \end{abstract}


\maketitle

        \section{Introduction\label{Introduction}}
                \TNSe{} is one of a handful of excitonic insulator (EI) candidates\cite{Takagi_EI,Wakisaka2012,KanekoTheory,BEC_in_TNSe,Takagi2016,Larkin2017,Werdehausen2018}, materials in which electrons and holes spontaneously bond to form a coherent condensate with a concomitant increase of the resistivity. Its quasi-one-dimensional electronic structure with a high  joint density of states for electron-hole pair excitations is a consequence of  parallel sets of Ta and Ni chains running along the $a$ axis of its crystal structure. Recent studies have scrutinized the electronic properties of \TNSe{} to reveal a phase transition of electronic origin associated with exciton condensation below $T_c$ = 326~K \cite{Takagi2016}, which is concomitant with an orthorhombic-to-monoclinic structural transition and apparently driven by electron-phonon coupling, in line with theoretical predictions\cite{KanekoTheory}.
                                The particular role that exciton-phonon coupling plays in \TNSe{} is very different from its role in indirect gap EI-candidates\cite{TmSeTe_Pressure_91,WachterBands,TiSe2_ARPES5}. For example, due to the indirect gap in $1T$-TiSe\ci{2}, momentum transfer from a phonon is necessary to create an exciton. For the same reason a charge density wave (CDW) appears, but the lattice is also subject to a $2 \times 2$ reconstruction which requires the presence of a charge-lattice interaction. \TNSe{}, on the other hand, is a direct gap material, so the emergence of an EI phase should not form a CDW and the structural transition, rather than forming a superlattice, is of a $q=0$ type. The direct band gap also allows excitons to be generated optically and phonons may instead screen the Coulomb interaction between electrons and holes and stabilize the excitons. In a recent optical-pump terahertz-probe study, a phonon-coupled state of the excitonic condensate  in \TNSe \ has been associated with the emergence of a  low frequency collective mode, which exhibits  combined properties of the coherent phonon and electron amplitude mode \cite{Werdehausen2018}. Exciton-phonon interactions have also been proposed to cause the emergence of doublets of exciton Fano resonances in \TNSe{} and the nearly isostructural \TNS{} at low temperatures and to cause a strong temperature dependence of the optical absorption peaks corresponding to interband transitions\cite{Larkin2017}. In monoclinic \TNSe{}, the exciton Fano resonances exhibit a giant spectral weight, which was attributed to antenna emission of overlapping exciton-phonon complexes. In orthorhombic \TNS{}, which does not exhibit a structural phase transition attributable to exciton condensation, the optical spectral weight of the exciton resonances is much smaller. This correspondence is consistent with the hypothesis that the low-temperature phase of \TNSe{} is an excitonic insulator.
                                
%                                Giant spectral weight of the exciton Fano resonances  attributed to antenna emission of overlapping exciton-phonon complexes in monoclinic \TNSe{} apparently distinguishes it from orthorhombic \TNS, which does not exhibit a structural phase transition associated with exciton condensation.  

				Motivated by these recent findings and to gain a better insight into the exciton-phonon coupling in these peculiar materials, we carried out a comparative study of the infrared phonon spectra of \TNSe{} and \TNS. We found that the position of the phonon modes and their polarization dependence in both compounds are in excellent agreement with lattice dynamics calculations. All phonons exhibit   rather conventional  softening with increasing temperature reflecting the  thermal expansion of the lattice due to anharmonicity. An enhanced damping of the $a$-axis phonon modes polarized along the Ta and Ni chains can be attributed to interaction with the electronic background, which gradually fills the energy gap of  \TNSe{} on approaching $T_c$, indicating near-zero-gap behavior at high temperature \cite{Takagi2016}. On the other hand, the electronic background  forms a broad nearly temperature-independent collective soft mode in the narrow-gap semiconductor \TNS{}. The electronic background behavior is consistent with the inference that the excitonic states in \TNSe{} are largely extended and overlapping along the chain direction, in accord with the EI\ hypothesis, but form local, spatially separated exciton-phonon bound states in \TNS{}.        
                
        \section{Experimental and computational details\label{Details}}
				Thin single crystals of \TNSe{} and \TNS{} with typical lateral dimensions 5~mm by 1\,--\,2~mm along the $a$- and $c$-axes, respectively, and thicknesses of 100\,--\,200~$\mu$m  were grown and characterised  as described elsewhere\cite{Takagi2016}, and cleaved prior to every optical measurement.

				The complex dielectric function was determined in the far-infrared (FIR, 4 -- 85~meV, 30 -- 700~cm$^{-1}$) and in a range of temperatures from 10~K to 350~K by combining near-normal incidence reflectivity and ellipsometry techniques.  The FIR Ellipsometric measurements  were performed  using a home-built  ellipsometer in combination with a Bruker IFS 66v/S FT-IR spectrometer at the IR1 beamline of the Karlsruhe Research Accelerator (KARA) at the Karlsruhe Institute of Technology. 

				The ellipsometric parameters $\Psi$ and $\Delta$ measured at angle of incidence 80\de{} (77.5\de{}) relative to the b-axis (normal to the $ac$-plane)\ of our samples define the complex ratio $r_p/r_s=\tan(\Psi)\exp(i\Delta)$, where $r_p$ and $r_s$ are the complex Fresnel coefficients for light polarized parallel and perpendicular to the plane of incidence, respectively. Sample orientations were selected with the $p$ component of the electric field vector to be parallel to the $ab$ ($cb$) plane. The pseudodielectric function  $\varepsilon^*(\omega) = \varepsilon^*_1 (\omega)+i\varepsilon^*_2 (\omega)=1+4\pi i[\sigma^*_1 (\omega)+i\sigma^*_2 (\omega)]/\omega$, derived by a direct inversion of $\Psi$ and $\Delta$ assuming bulk isotropic behavior of the sample, provides the $a$- (or $c$-) axis tensor element of the dielectric tensor $\varepsilon_a$ (or $\varepsilon_c$) with a minor contribution of the orthogonal components of the dielectric tensor $\varepsilon_b$ and $\varepsilon_c$ (or $\varepsilon_a$).

				Terahertz (THz) spectra of \TNSe{} were obtained in the range 0.35\,--\,1.75~THz (1.5\,--\,7.2~meV) by high-resolution time-domain THz spectroscopy in the transmission configuration. The spectra were measured with a LaserQuantum HASSP-THz spectrometer using the high-speed asynchronous optical sampling technique (ASOPS) at a 1~GHz repetition rate. The complex dielectric function was calculated via inversion of the Fresnel transmission coefficients for an electric field waveform polarized parallel to the $a$ or $c$ axis  transmitted through a 1.3 mm diameter aperture, with an empty aperture used as a reference. 

                We also measured the low-temperature near normal FIR reflectivity  with polarized light using a  Bruker Vertex 80v Fourier transform spectrometer. An \textit{in situ} gold evaporation stage was used to coat the \TNSe{} sample with gold for reference measurements. Ellipsometric data for the corresponding components of the dielectric tensor  above the FIR energy range obtained in Ref.~\onlinecite{Larkin2017} were used for the high-energy extrapolation and the subsequent Kramers-Kronig (KK) transformation of the reflectivity data. The low-energy extrapolation of the reflectivity data assumed an insulating sample and was taken as tending to a constant value.

                \begin{table}
                	\begin{ruledtabular}
                    	\begin{tabular}{ccccc}
                        	\multicolumn{5}{c}{\TNSe{}, $C2/c$ (15), $\beta=90.53^\circ{}$} \\
                        	\multicolumn{2}{l}{Latt. const. (\small\AA)} & $a=3.496$ & $b=12.829$ & $c=15.641$ \\
                        	\hline 
                            Atom    & Site  & X             & Y               & Z                     \\
                            \hline
                            Ta (1)  & 8f    & -0.00793      &\ 0.22135        &\ 0.11044 \\
                            Ni (1)  & 4e    &\ 0.00000      & -0.29890        &\ 0.25000 \\
                            Se (1)  & 8f    &\ 0.00530      & -0.41970        &\ 0.13800 \\
                            Se (2)  & 8f    & -0.00510      &\ 0.14560        & -0.04910 \\
                            Se (3)  & 4e    &\ 0.00000      &\ 0.32710        &\ 0.25000 \\
                            \hline\hline
                            \multicolumn{5}{c}{\TNS{}, $Cmcm$ (63)}                   \\
                            \multicolumn{2}{l}{Latt. const. (\small\AA)} & $a=3.415$ & $b=12.146$ & $c=15.097$ \\
                            \hline
                            Atom    & Site  & X             & Y               & Z                     \\
                            \hline
                            Ta (1)  & 8f    &  0 &\ 0.22082         &\ 0.10879 \\
                            Ni (1)  & 4c    &  0 &\ 0.69692         &\ 0.25000 \\
                            S  (1)  & 8f    &  0 & -0.41718         &\ 0.13527 \\
                            S  (2)  & 8f    &  0 &\ 0.14850         & -0.05026 \\
                            S  (3)  & 4c    &  0 &\ 0.31968         &\ 0.25000 \\
                        \end{tabular}
                    \end{ruledtabular}
                    \caption{Structural details of \TNSe{} and \TNS{}, as obtained by Sunshine and Ibers\cite{SunshineIbers1985}.\label{Crystal_Data}}
                \end{table}
                \begin{figure*}
                    \includegraphics[width=166mm]{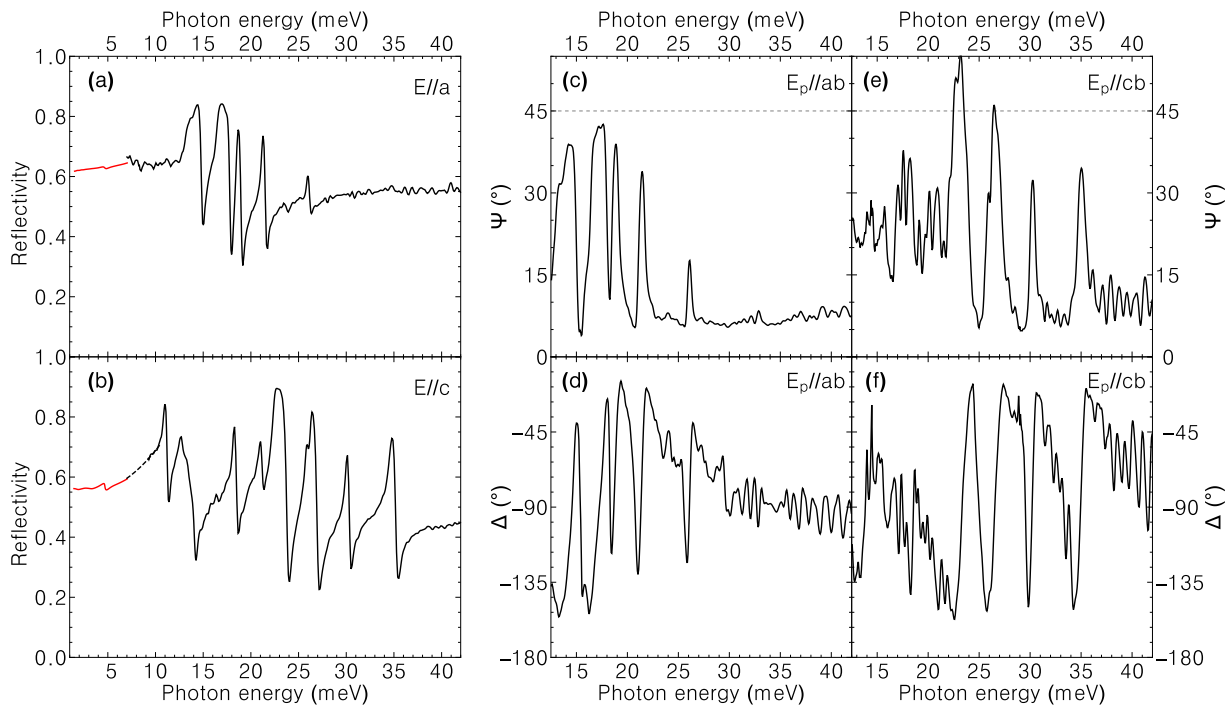}
                    \caption{Experimental results at 10~K of (a,b) the normal-incidence reflectivity for light polarized along the $a$- and $c$-axes, and reflectivity in the THz range, as calculated from the results of transmission measurements (red lines); (c,d) ellipsometric angles $\Psi$ and $\Delta$ measured at an 80\de{} angle of incidence with the $a$-axis in the plane of incidence; (e,f) the same for the $c$-axis and a 77.5\de{} angle of incidence.\label{Raw_Data_10K}}
                \end{figure*}
                
                For the calculations we have employed density functional perturbation theory\cite{baroni_dfpt} as implemented in \textsc{quantum espresso}\cite{qe} to calculate the vibrational modes of both compounds at the $\Gamma$-point. We used the experimental structural data listed in Table~\ref{Crystal_Data} according to Ref.~\onlinecite{SunshineIbers1985}, projector augmented plane-wave\cite{paw_bloechl} pseudopotentials\cite{pslibrary, pseudos} and the local density approximation. We set the energy cutoff for the wave-functions (charge-density) to 60~Ry (600~Ry). The initial structure was optimized to reduce the stress below 0.1~kBar and the residual forces below 0.1~mRy/Bohr per atom prior to the lattice dynamics calculation.
                
                \begin{table}
                    \begin{ruledtabular}
                        \begin{tabular}{ccccccc}
                            \multicolumn{4}{c}{Experiment} & \multicolumn{3}{c}{Calculation} \\
                            \cline{1-4} \cline{5-7}
                            $N$ & $\omega_0$ (meV) & $\Delta\varepsilon$ & $\gamma$ (meV) & $N$ & $\omega_0$ (meV) & Ch. \\
                            \hline
                        	\multicolumn{4}{c}{$a$-axis}& \multicolumn{3}{c}{$B_u$} \\
                        	 1 &  4.71 &  0.55 & 0.37	&  1 &   4.37 & $B_{1u}$ \\
                        	 2 & 14.00 & 10.92 & 0.62	&  2 &  11.65 & $B_{3u}$ \\	
                             4 & 16.74 &  6.21 & 0.40	&  3 &  12.89 & $B_{1u}$ \\
                             6 & 18.61 &  0.95 & 0.20	&  4 &  17.40 & $B_{3u}$ \\
                             5 & 21.26 &  0.93 & 0.20	&  5 &  19.27 & $B_{1u}$ \\
                             9 & 26.10 &  0.28 & 0.29	&  6 &  20.51 & $B_{3u}$ \\
                                               & & & 	&  7 &  22.25 & $B_{1u}$ \\
                        	\multicolumn{4}{c}{$c$-axis}&  8 &  26.13 & $B_{1u}$ \\
							 1 &  4.71 &  0.55 & 0.37	&  9 &  28.01 & $B_{3u}$ \\
                             3 & 10.88 &  6.90 & 0.31   & 10 &  29.27 & $B_{1u}$ \\
                             2 & 12.67 &  8.32 & 1.13   & 11 &  35.64 & $B_{1u}$ \\
                             6 & 18.24 &  1.56 & 0.22 \\    
                             5 & 21.05 &  1.14 & 0.27 \\      
                             7 & 22.49 &  4.39 & 0.25 \\
                             8 & 25.89 &  0.71 & 0.26 \\
                             9 & 26.22 &  0.95 & 0.33 \\
                            10 & 30.05 &  0.54 & 0.29 \\
                            11 & 34.63 &  0.96 & 0.45
                                \end{tabular}
                        \end{ruledtabular}
                        \caption{The observed phonon modes in \TNSe{} for the two polarizations and the calculated eigenfrequencies of the 11 $B_u$ modes along with a possible assignment. The last column indicates the dominant character of the modes according to orthorhombic symmetry.\label{Modes}}
                \end{table}
                
	        \section{Phonon spectrum}
                Figure~\ref{Raw_Data_10K} shows the low-temperature reflectivity and ellipsometric angles of \TNSe{} measured in configurations for $E_p||ab$ and $E_p||cb$. Interference fringes due to multiple reflections are particularly noticeable in the $\Delta$ curves, indicating good transparency of a thin, flat sample. The absence of an underlying electronic background indicates the high purity of the samples. The non-physical values of $\Psi > 45^\circ{}$ are related to the contribution of the orthogonal components of the dielectric tensor\cite{Schubert2000,Humlicek2000}. 

\begin{figure}
                    \includegraphics[width=70mm]{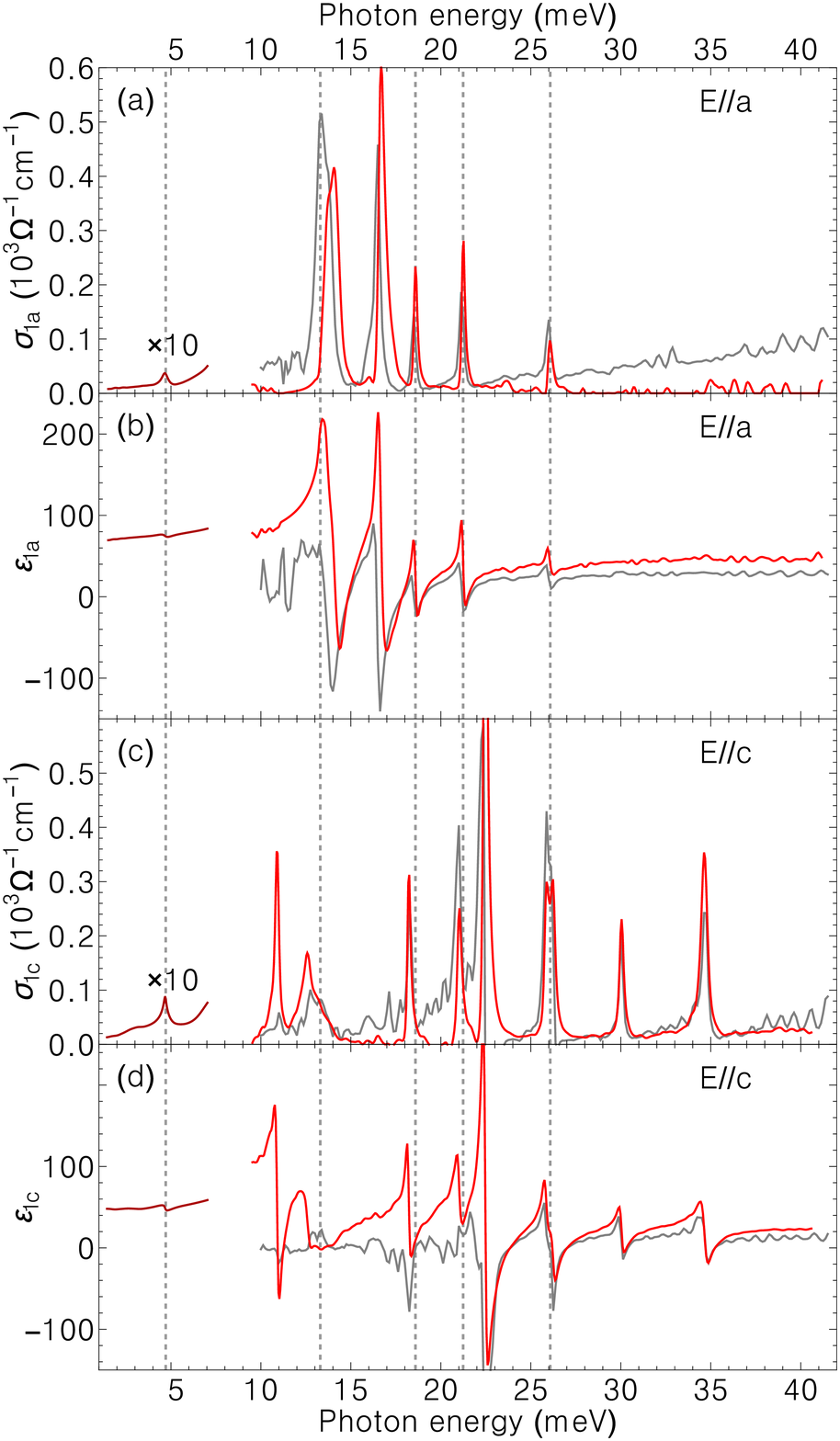}
                    \caption{(a) Real part of the $a$-axis optical conductivity, as obtained from a KK-transformation of the reflectivity spectra (red lines), from THz transmission (magnified by a factor of 10, maroon lines) and from ellipsometry (gray lines). (b) The corresponding dielectric permittivity. (c,d) Same as (a,b), but for the $c$-axis. The gridlines highlight five $a$-axis modes which have close-lying counterparts in the $c$-axis spectrum.\label{Epsilon_KK}}
                \end{figure}

				Figure~\ref{Epsilon_KK} shows the low-temperature $a$- and $c$-axis optical conductivities and dielectric permittivities, as extracted from the reflectivity and THz transmission spectra, and directly compares them to the ellipsometric results. The phonon spectra obtained by direct inversion of the ellipsometric angles with respect to the pseudodielectric function are in good agreement with reflectivity data, especially at higher energies, where diffraction effects due to the finite sample size are less of a limitation. These data confirm that we are primarily observing the in-plane contribution of the dielectric tensor, despite the strong anisotropy of \TNSe{}. The $a$-axis spectrum shows 6 main peaks with some internal structure, while at least 10 peaks are found in the $c$-axis, along with a collection of less-prominent features.

                \begin{table}
                	\begin{ruledtabular}
                    	\begin{tabular}{cccc}
                        	\multicolumn{3}{c}{Experiment} & Calculation \\
                            \cline{1-3} \cline{4-4}
                            $\omega_0$ (meV) & $\Delta\varepsilon$ & $\gamma$ (meV) & $\omega_0$ (meV) \\
                            \hline
	                        \multicolumn{3}{c}{$a$-axis}& $B_{3u}$ \\
                            (19)  &      &          & 16.35 \\
                            (23)  &      &          & 19.04 \\
                            32.45 & 0.04 & 0.18     & 33.61 \\
                            35.67 & 0.23 & 0.33     & 38.21 \\
                            \\
	                        \multicolumn{3}{c}{$c$-axis}& $B_{1u}$  \\
                                       &      &    &  4.85 \\
							21.22 & 4.21 & 0.36    & 19.42 \\
							29.02 & 3.25 & 0.73    & 28.99 \\
							{}    & {}   & {}      & 36.75 \\  
							37.77 & 0.83 & 0.24    & 37.89 \\
							43.65 & 0.06 & 0.12    & 44.18 \\
							      &      &         & 49.66 \\
                        \end{tabular}
                    \end{ruledtabular}
                    \caption{The observed phonon modes in \TNS{} for the two polarizations and the calculated eigenfrequencies of the 4~$B_{3u}$
and 7~$B_{1u}$ modes along with a possible assignment.\label{Modes_S}}
                \end{table}
                \begin{figure*}
                    \includegraphics[width=160mm]{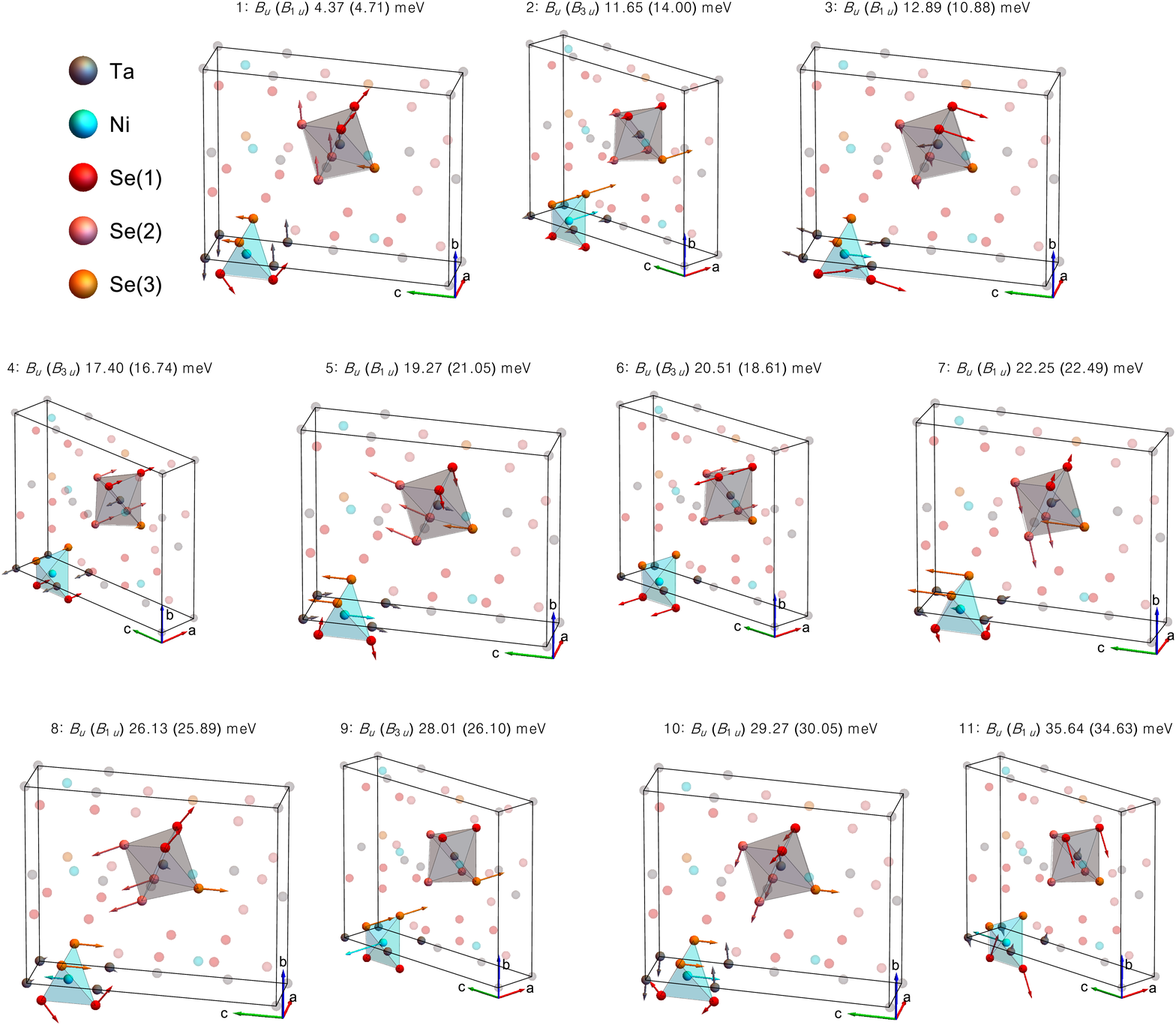}
                    \caption{The calculated displacement patterns of the 11~$B_u$ optical phonon modes. The black rhomboid shows the boundary of a single unit cell. Arrows show the directions and relative sizes of the displacements for each phonon mode. The labels show the mode number (as in Table~\ref{Modes}), the mode character ($B_u$ for all modes), the mode character in orthorhombic symmetry (in brackets), the calculated eigenfrequency and the experimentally observed mode energy (as seen in the $a$ ($c$) axis for $B_{3u}$ ($B_{1u}$) modes; in brackets). Every Ni atom (cyan) is tetrahedrally coordinated by Se atoms (red, pink, orange, with different shades corresponding to nonequivalent atomic sites, in particular, orange corresponds to the $4e$ Wyckoff position) and a rectangular plaquette of Ta atoms (gray); this coordination is highlighted by a cyan tetrahedron. Likewise, every Ta atom is octahedrally coordinated by Se atoms (gray octahedron). The displacement pattern and relative positions of the second Ni atom of the primitive cell and its surrounding atoms are obtained by a spatial inversion and a phase reversal of the shown oscillations. Similarly, for the remaining three Ta atoms the positions and displacements are found by an inversion in $x$ and $z$ ($x \parallel a,\,z \parallel c$) with a phase reversal, a $y$ inversion ($y \perp x,\, y \perp z$), or a complete spatial inversion and a phase reversal.\label{Phonon_displacement}}
                \end{figure*}
                \begin{figure}
                    \includegraphics[width=80mm]{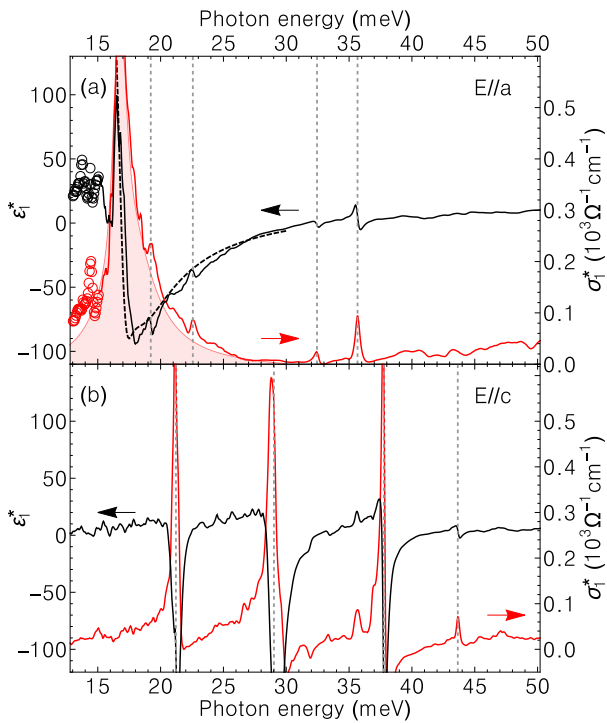}
                    \caption{Results of ellipsometric measurements of \TNS{}. (a) Real parts of the $a$-axis optical conductivity and dielectric permittivity of \TNS{} averaged over $T=10-150\,\mathrm{K}$. The shaded area shows an unusually large soft absorption band in the optical conductivity and the black dashed line is its associated counterpart in the permittivity. (b) Real parts of the $c$-axis optical conductivity and dielectric permittivity of \TNS{} at $T=10\,\mathrm{K}$. Vertical gridlines show the positions of the phonon peaks.\label{TNS_avg_elli}}
                \end{figure}
                \begin{figure*}
                    \includegraphics[width=177mm]{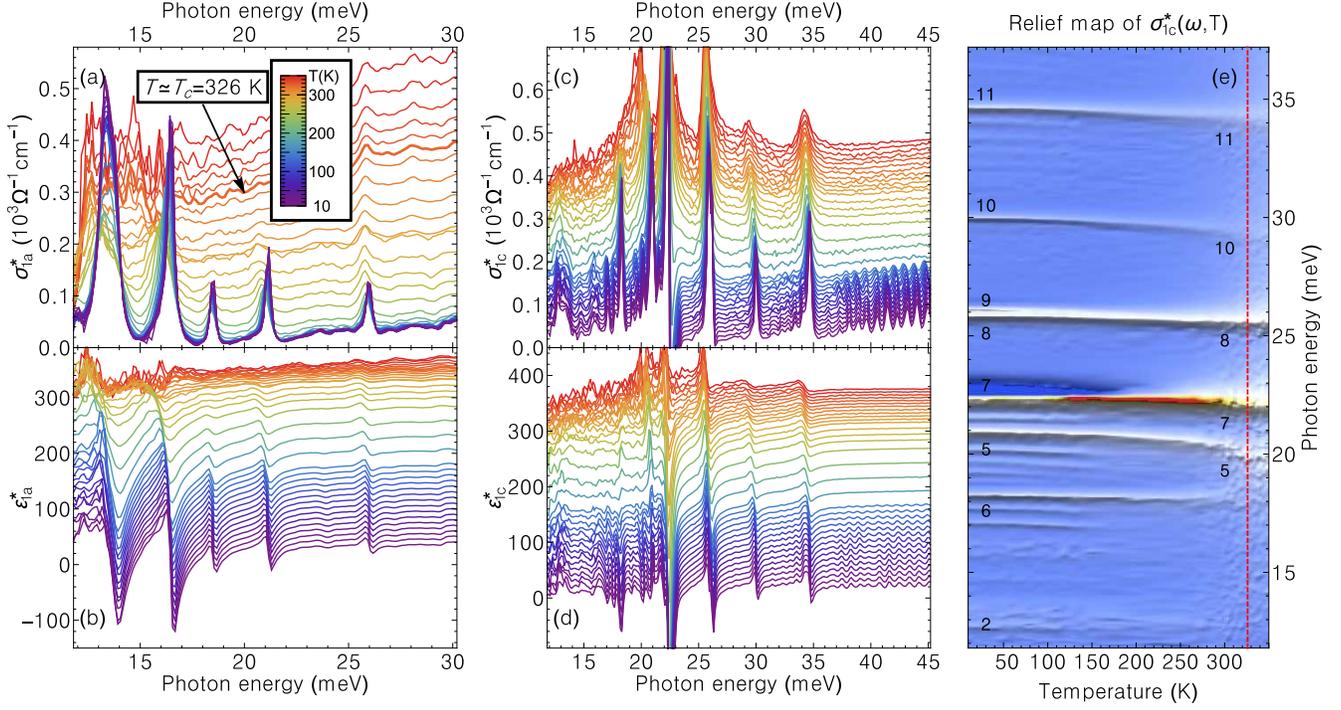}
                    \caption{Full set of the temperature dependent ellipsometric measurements of \TNSe{}. (a,b) Real parts of the optical conductivity and dielectric permittivity along the $a$-axis in \TNSe{}, respectively. The permittivity spectra are shown with offsets equal to $T/\mathrm{K}$. Note that the optical conductivity is shown \textit{without} offsets. The thick line is the measurement at $T=325\,\mathrm{K}$, the closest temperature to $T_c=326\,\mathrm{K}$. (c,d) The same for the $c$-axis. Optical conductivity (permittivity) is shown with an offset of $T \cdot \Omega^{-1} \mathrm{cm}^{-1}\! / \mathrm{K}$ ($T/\mathrm{K}$). (e) Relief map of the $c$-axis optical conductivity, illustrating the softening and broadening of the phonon modes. Numbers on the left-hand-side indicate the phonon modes (as listed in Table~\ref{Modes}) observed at the lowest temperature, while numbers on the right-hand-side show modes which persist at the highest temperature. The vertical line marks the structural transition temperature.\label{TNSe_a_elli}}
                \end{figure*}
                \begin{figure}
                    \includegraphics[width=80mm]{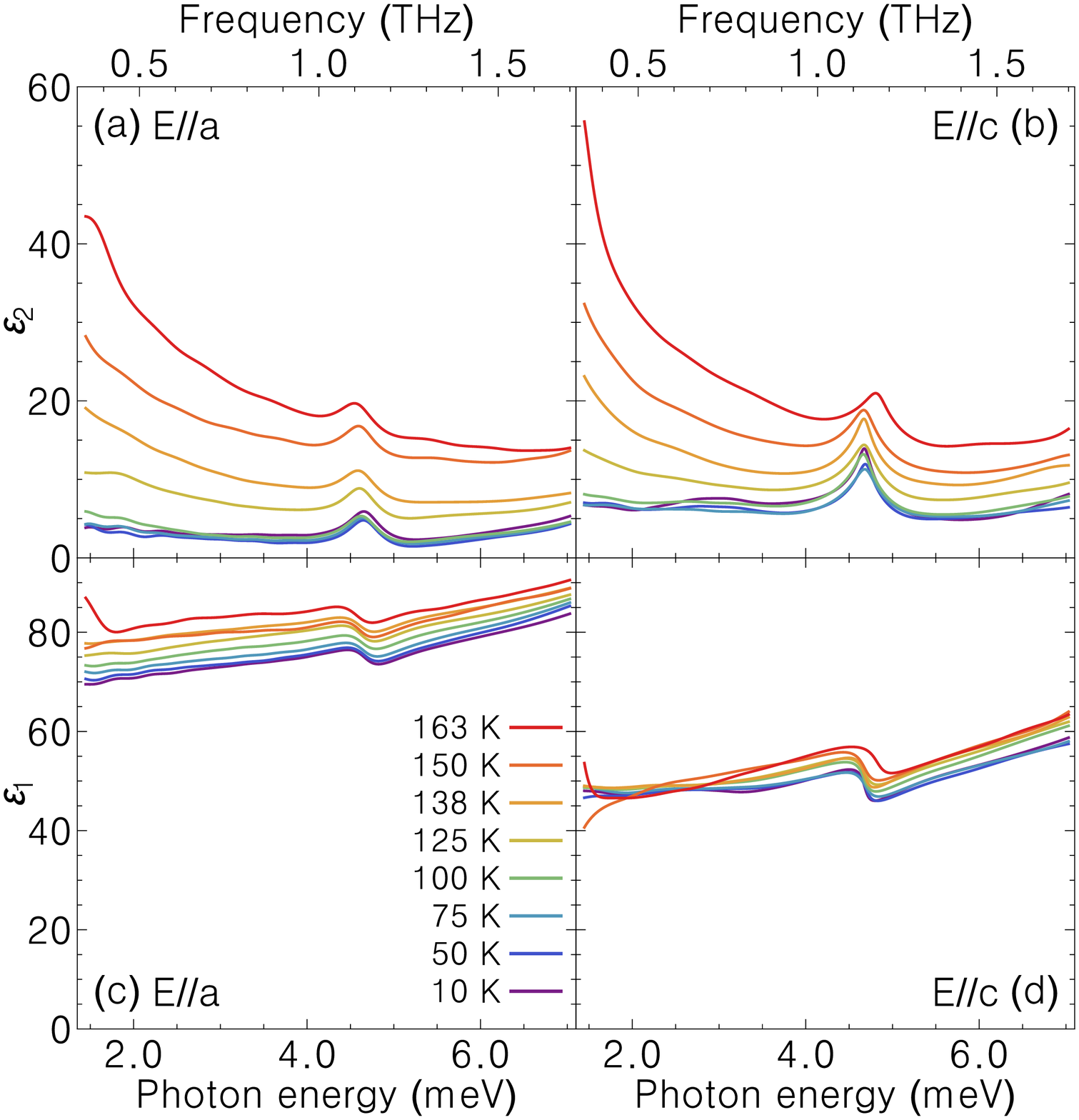}
                    \caption{The temperature-dependent set of THz spectra for \TNSe{} measured by time-domain THz spectroscopy. (a, c) The imaginary part of the dielectric permittivity as measured along the a- and c-axis, respectively. (b, d) The same for the real part of the dielectric permittivity.\label{TNSe_a_THz}}
                \end{figure}

                The symmetry of the lattice admits up to 21 dipole-active phonon modes (for brevity, we omit the acoustic modes in the entire discussion), of which 11 $B_u$ modes can induce a dipole moment along the $a$- and $c$-axes, while the remaining 10 $A_u$ modes induce a moment along the $b$-axis. The structure of \TNSe{} is very close to the higher-symmetry structure of \TNS{} where the 11~$B_u$ modes separate into 7~$B_{1u}$ and 4~$B_{3u}$ modes, with dipole activity along the $c$- and $a$-axis, respectively. Accordingly, of the 11~$B_u$ modes of \TNSe{}, the character of 4 is very close to a $B_{3u}$-type, while the other 7 are of a $B_{1u}$-type, as shown in Table~\ref{Modes}. However, this minor lowering of the symmetry already leads to noticeable dipole activity in both polarizations, as was confirmed by DFT calculations. Since a combined total of 16 peaks is found in the $a$- and $c$-axes, at least five $B_u$ modes must be simultaneously active in both polarizations. To explore this possibility we attempted to simultaneously fit the spectra for both polarizations with a set of no more than 11 complex Lorentzian oscillators, attempting to capture only the most prominent features. Indeed, a subset of the $a$-axis phonons has counterparts in the $c$-axis spectrum with closely matching frequencies. The discrepancies between the observed energies in the signatures of what is presumably the same mode may be partially explained by the conservative fitting procedure which was aimed at using a minimal set of Lorentzians, rather than fitting the spectra very accurately.
                
                DFT calculations yield the eigenfrequencies $\omega_0$ of the 11 $B_u$ modes and their atomic displacement patterns. Their results are listed in Table~\ref{Modes} and shown in Fig.~\ref{Phonon_displacement}. Since \TNSe{} is metallic in DFT, the Born charge tensor is not accessible and we are forced to revert to the zeroth-order approximation, i.e. that the dipole moment is collinear with the eigendisplacement. Even under these restrictive assumptions, it follows from the calculations that all $B_u$ modes tending to a $B_{1u}$ ($B_{3u}$) character might to a lesser or greater extent exhibit some activity also along the $a$ ($c$) axis.
                
                \begin{figure*}
                    \includegraphics[width=177mm]{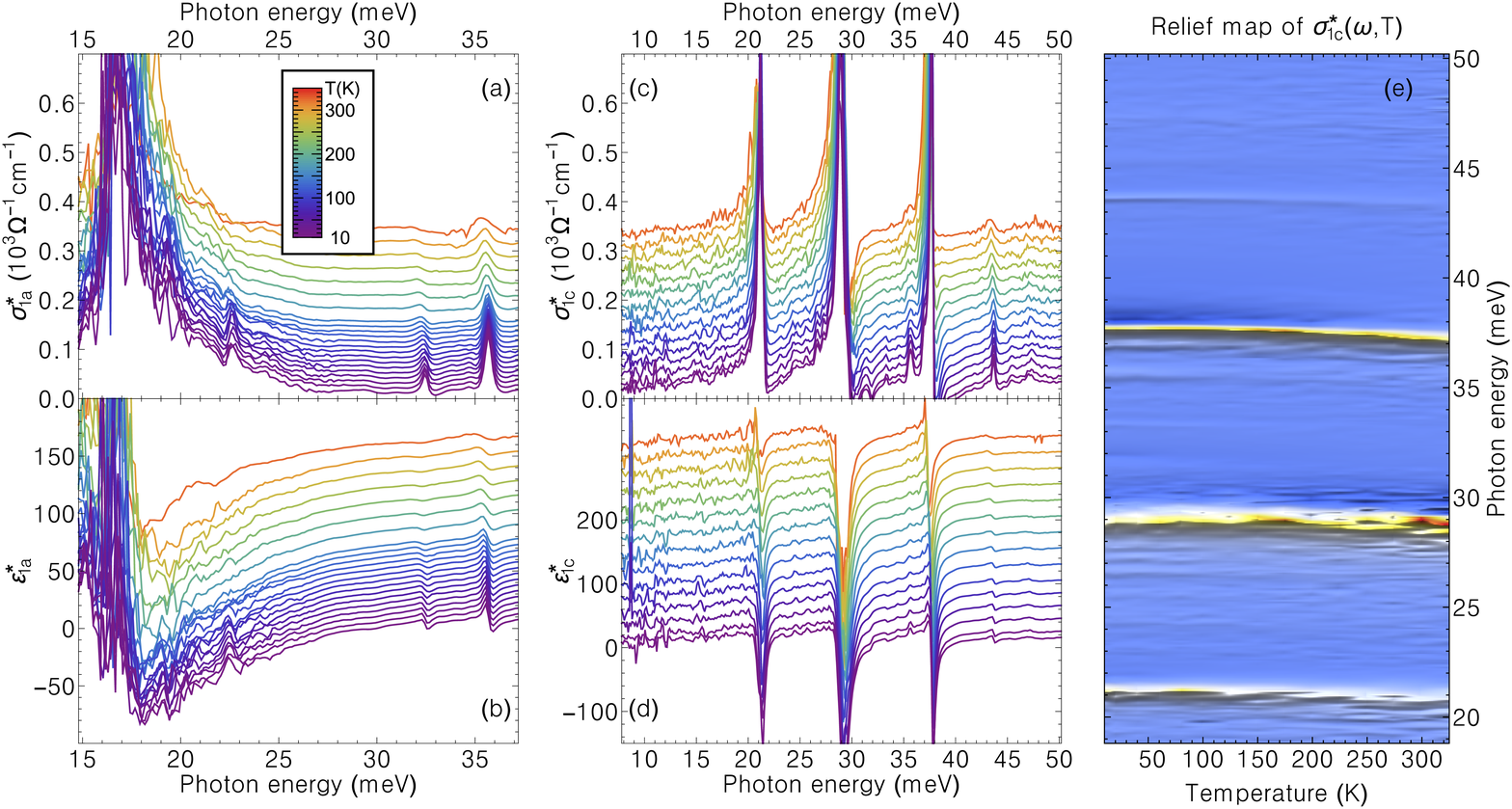}
                    \caption{Full set of the temperature dependent ellipsometric measurements of \TNS{}. (a,b) Real parts of the optical conductivity and dielectric permittivity along the $a$-axis in \TNS{}, respectively. Optical conductivity (permittivity) is shown with an offset of $T \cdot \Omega^{-1} \mathrm{cm}^{-1}\! / \mathrm{K}$ ($0.5 \cdot T/\mathrm{K}$). (c,d) The same for the $c$-axis. Optical conductivity (permittivity) is shown with an offset of $T \cdot \Omega^{-1} \mathrm{cm}^{-1}\! / \mathrm{K}$ ($T/\mathrm{K}$). (e) Relief map of the $c$-axis optical conductivity, illustrating the softening and broadening of the phonon modes.\label{TNS_a_elli}}
                \end{figure*}
                
                Based on the calculated displacement patterns, the expected polarizations, and the observed and calculated eigenfrequencies of the phonon modes, we were able to perform an assignment. The 8 highest energy modes agree very well with the calculations, with 3 of them showing activity in both polarizations (gridlines in Fig.~\ref{Epsilon_KK}, repeated numbers in the first column of Table~\ref{Modes}). The order of modes 5 and 6 is inverted, since mode 5 is of a predominantly $B_{1u}$ character, however, of the two candidate modes, the one higher in energy survives above the temperature of the structural transition in the $c$-axis spectra of \TNSe{} (see the end of this section for details). The THz peak unambiguously corresponds to the lowest energy phonon mode. Since the $c$-axis peak at 10.88~meV is not seen in the $a$-axis, we find it likely to correspond to mode No.~3, rather than No.~2. The remaining mode No.~2 apparently manifests itself in both axes as a rather irregularly shaped peak with a discrepancy between its frequencies in the two polarizations (14.00~meV in $a$ vs. 12.67~meV in $c$). As this mode is close to the edge of the measured range, the KK transformation is sensitive to uncertainties in the reflection coefficient seen in the experimental noise between 7 and 13~meV in the $a$-axis reflectivity data, which may be a cause of this mismatch. Indeed, the agreement between the two polarizations in the ellipsometric data is much better.
                
                \TNS{} is also metallic in DFT, hence the IR cross-sections also cannot be directly calculated. However, it is in a higher symmetry group than \TNSe{}. Here the 11 $B_u$ modes separate into 4 $B_{3u}$ modes active only along the $a$-axis and 7 $B_{1u}$ modes for the $c$-axis. The 10 $A_u$ modes separate into 7 $B_{2u}$ modes with dipole activity in the $b$-axis, while the remaining 3 $A_u$ modes become silent, and are neither dipole-, nor Raman-active. The separation of all phonon modes along separate crystallographic axes ensures a diagonalized dielectric tensor which, in turn, leads to less-frequently occuring negative imaginary parts of the pseudodielectric function. The calculated modes of \TNS{} are listed in Table~\ref{Modes_S} and the direction of dipole activity is immediately available from the irreducible representations of the modes. The experimental results are presented in Fig.~\ref{TNS_a_elli}. The assignment for \TNS{} is much simpler. The 4 phonon lines which were unambiguously observed in the $c$-axis response are in very good agreement with the calculated eigenfrequencies. The two higher-energy modes of the $a$-axis are also clearly visible and agree well with calculations. The remaining two phonon modes are strongly masked by noise in the raw data. However, by taking the mean of the pseudodielectric functions obtained at temperatures from 10~K to 150~K (Fig.~\ref{TNS_avg_elli}), we find two likely candidates which persist in the spectra and are roughly KK-consistent in their real and imaginary parts. Additionally, in the $c$-axis optical conductivity we see a weak peak near 47~meV with a KK-consistent counterpart in the permittivity which may correspond to the calculated mode with energy 49.79~meV. The mode with calculated energy 36.89~meV might be represented by the low-energy shoulder of the peak at 37.77~meV. The last mode at 4.92~meV, without a possible assignment, is certainly beyond the experimentally-accessible spectral range. The most prominent peaks are fitted with a set of Lorentzians and presented in Table~\ref{Modes_S} with possible assignments given by corresponding rows for the experimental data and the calculation.
                
                Since \TNS{} and \TNSe{} are isoelectronic and nearly isostructural, it may be expected that the differences between their phonon spectra are largely determined by the mass difference of the chalcogenides. As is seen from the displacement patterns in Fig.~\ref{Phonon_displacement}, modes 6\,--\,8 are mostly vibrations of the chalcogenide, and indeed, the corresponding phonon modes of \TNS{} are found close to $\sqrt{m_{Se}/m_{S}}\approx 1.57$ times higher in energy than in \TNSe{}.
                
				In this way, we find the phonon modes in \TNSe{} and \TNS{} to be in good agreement with the DFT calculations and earlier observations. The phonon spectra of the two materials are similar, as one could expect from the similarity of their crystal structures. The somewhat higher energies of the vibrational modes in \TNS{} are well aligned with the mass difference of sulphur and selenium. As observed earlier\cite{Larkin2017}, the interband transitions of \TNSe{} (\TNS{}) are coupled to phonon modes with energies in the range of 13\,--\,17~meV (21\,--\,24~meV); this is in good agreement with the presented eigenfrequencies (both calculated and measured) of the phonons.

				To characterize the temperature dependences of the lattice vibrations we performed ellipsometric measurements of \TNSe{} and \TNS{} for temperatures ranging from 10~K to 350~K (Figs.~\ref{TNSe_a_elli}~and~\ref{TNS_a_elli}). The phonon modes demonstrate a rather generic manifestation of anharmonicity with an expected softening and broadening of the lineshapes with increasing temperature. This temperature dependence is readily seen in the relief maps for the $c$-axis optical conductivities of \TNSe{} and \TNS{} (Figs.~\ref{TNSe_a_elli}e~and~\ref{TNS_a_elli}e). Furthermore, Fig.~\ref{TNSe_a_elli}e shows the disappearance of three modes in \TNSe{} above the structural transition temperature $T_c = 326\,\mathrm{K}$, which allows us to unambiguously assign them the $B_{3u}$ character, i.e. $B_u$ modes 2, 6, and 9, as numbered in Table~\ref{Modes}. This behavior reflects the fact that the monoclinic distortion permits the $a$-axis polarized modes to be active also in the $c$-direction. In \TNS{}, on the other hand, there is no long-range monoclinic distortion and therefore no change in the number of modes, as can be seen in Fig.~\ref{TNS_a_elli}e.

				The enhancement of phonon damping  makes  the $a$-axis phonon modes polarized along the Ta and Ni chains (Figs.~\ref{TNSe_a_elli}a~and~\ref{TNS_a_elli}a) poorly discernible at high temperatures as they become superimposed on the electronic background. 

\section{Electronic background}

				The electronic background rapidly  fills the optical band gap $E^{op}_g\approx0.16\,\mathrm{eV}$ of \TNSe{} on approaching the monoclinic-to-orthorhombic phase transition at $T_c = 326\,\mathrm{K}$ (note Fig.~\ref{TNSe_a_elli}(a)) where the optical conductivity is shown without offsets) and forms a broad infrared continuum peaked at $\sim 0.35\,\mathrm{eV}$ that bears resemblance to small polaron behavior\cite{Larkin2017}. The associated spectral weight is transferred from the above-gap states and remains nearly conserved in the infrared below $\sim 1.2\,\mathrm{eV}$ (Fig.~2(c) in Ref.~\onlinecite{Larkin2017}). According to the interpretation of specific-heat measurements, most of the entropy change associated with the transition originates from the electronic entropy indicating an electronically driven phase transition, presumably due to exciton condensation below $ T_c$\cite{Takagi2016}. Results of dc electrical resistivity measurements parallel to the Ta and Ni chains suggest that the high-temperature orthorhombic phase of \TNSe{} is a near zero-gap semiconductor\cite{Takagi2016}. Indeed, our FIR ellipsometry data measured up to 350~K ($T>T_c$) show positive dielectric permittivity $\varepsilon_1(\omega)$ down to the low-frequency cutoff at $\sim 10\,\mathrm{meV}$, and thus confirm near-zero electron density of states at the Fermi level. 

				The development of the electronic background at phonon frequencies can be also analyzed on the basis of transmission THz time-domain spectroscopy data. The applicability of this technique  is limited to temperatures below 165~K for which the 170~$\mu$m thick \TNSe{} sample remains transparent. Fig.~\ref{TNSe_a_THz} shows the temperature dependent real and imaginary parts of the THz (0.35\,--\,1.75~THz, 1.5\,--\,7.2~meV) dielectric function measured along $a$- and $c$-axes of \TNSe{}. As the temperature increases above 100~K, a Drude-like upturn in $\varepsilon_2(\omega)$ toward the origin develops in both polarizations that evidences coherent charge transport both along and across the chains. The Drude intraband spectral weight strongly increases as a function of temperature above 100~K, which implies its origin from an increasing  population of thermally activated electron and hole pairs in narrow bands of energy near the bottom of the conduction band and the top of the valence band, respectively. Based on the fit to both $\varepsilon_1(\omega)$ and $\varepsilon_2(\omega)$ measured at 163~K ($k_BT\approx 14\,\mathrm{meV}$), we estimated the Drude fit parameters, i.e. the plasma frequency $\omega_p$ and scattering rate $1/\tau_D$, to be approximately 20~meV and 8~meV, respectively. We attribute these features to a small density of free carriers whose spectral weight translates into an effective number of about $2.1 \cdot 10^{-4}$ electrons per Ni atom, more than three orders of magnitude less than the total spectral weight, $N^{IR}_{eff}\sim 0.7\,e/\mathrm{Ni}$, of the low energy transitions across the gap between the Se~$4p$/Ni~$3d$ hole and Ta~$5d$ electron bands of \TNSe{} ($E_G \approx 160\,\mathrm{meV}$ at $T=160\,\mathrm{K}$)\cite{Larkin2017}. The associated changes in $\varepsilon_2(\omega)$ observed for both polarizations are consistent with a moderate in-plane anisotropy in $dc$ resistivity, which is supposed to vanish in the EI state due to the associated flattening of the one-dimensional dispersion\cite{Takagi2016}. Significant anisotropy in the real part of the THz dielectric function $\varepsilon_1(\omega)$ is caused by higher density of states for direct electron-hole excitations along the Ta\,--\,Ni zigzag chains and reflects the high polarizability of the Ta\,--\,Ni bonds. The associated interband transitions form a broad maximum at about 0.4 eV in the low-temperature mid-infrared conductivity spectra and make a dominant contribution to the static dielectric permittivity $\varepsilon_1(\omega\rightarrow0)$. The continuous upward shift of the  $\varepsilon_1(\omega)$ curves with increasing temperature in Fig.~\ref{TNSe_a_THz}b corresponds to the marked temperature-induced redshift and broadening of the mid-infrared band shown in Figs.~2(e,f) of Ref.~\onlinecite{Larkin2017}. Therefore, the filling of the gap region can be understood in terms of the development of a thermally activated coherent Drude peak and due to the massive transfer of the spectral weight from the above-gap states on approaching $T_c$, implying the  predominant incoherent hopping of charge carriers along the Ta\,--\,Ni chains in the orthorhombic phase of \TNSe{}.
     
				Compared with the effective zero gap in the orthorhombic \TNSe{}, the optical band gap in the isostructural \TNS{} remains at 0.2\,--\,0.3~eV over the entire range of measured temperatures. Since the gap value is supposedly larger than the exciton binding energy, \TNS{} does not show any signature of an EI-like transition and remains orthorhombic down to 2~K\cite{Takagi2016}. Notwithstanding the relatively large optical gap, the $B_{3u}$ phonon modes polarized along the $a$-axis of \TNS{} are superimposed on an in-gap soft electronic mode peaked near 16~meV, which is highlighted by the shaded area in Fig.~\ref{TNS_avg_elli}(a). This mode appears to be a feature that distinguishes between \TNSe{} and \TNS{}: while the energy gap of \TNSe{} has no in-gap states and is rapidly filled on approaching $T_c$, in \TNS{} instead an in-gap absorption peak persists with little to no temperature dependence. Its signatures are clearly present in both the optical conductivity and dielectric permittivity and are KK-consistent with one another, which allows us to accurately measure its spectral weight to be equal to $1.4 \cdot 10^{-3}\,e/\mathrm{Ni}$. This is more than two orders of magnitude less than the total spectral weight of the electron-hole transitions across the direct gap between the S~$3p$/Ni~$3d$ and Ta~$5d$  bands, $N^{IR}_{eff}\sim 0.34\,e/\mathrm{Ni}$, while comparable with the effective number of electrons associated with the quantum interference effect illustrated in Figs.~3(a-d) of Ref.~\onlinecite{Larkin2017}, accounting for 12\,--\,15\% of the spectral weight transferred across the exciton-Fano resonance frequency. The spectral weight of the excitonic resonances observed in both \TNS{} and \TNSe{} and their strong Fano interference with continuum excitations indicate that the exciton states are self-localized via exciton dressing with a cloud of phonons. The associated local deformation of the crystal lattice around the exciton can render two neighboring Ta\,--\,Ni bonds nonequivalent. In \TNS{} the exciton-phonon complexes are sparse and do not develop a long-range order. This local violation of symmetry allows the excitations to have a nonzero energy, which we see in the dielectric function as a soft localized excitation. In \TNSe{}, on the other hand, the giant spectral weight of the exciton Fano resonances due to antenna emission of large exciton-phonon complexes implies that they are strongly overlapping and probably span the entire crystal, hence preserving the translational symmetry of the lattice. Due to this symmetry the internal excitations of the exciton-polaron aggregates  may support propagating low-energy excitations akin to acoustic phonon modes.

\section{Conclusions}
	In summary, applying several spectroscopic techniques, we were able to measure the in-plane FIR spectra of \TNSe{} and \TNS{}. The obtained phonon modes are in excellent agreement with DFT lattice dynamics calculations, which allowed us to perform a mode assignment. We find that the monoclinic distortion presumably associated with an establishment of the EI state in \TNSe{} allows bidirectional $ac$ plane activity of the $B_{1u}/B_{3u}$ modes, as they are assigned in the orthorhombic structure of \TNS{}. The two compounds demonstrate distinct behaviors of the FIR electronic background: the optical gap in \TNSe{} fills with increasing temperature both from the low-energy side by the appearance of thermally activated charge carriers, and from the high-energy side due to the softening of the first interband transition. Instead, in \TNS{}, a persistent and temperature-independent electronic mode is present within the gap at around 16~meV. This difference reflects the different nature of the ground states of these two compounds, where extended and overlapping exciton-phonon complexes in \TNSe{} lead to a lowering of the crystal symmetry, whereas sparse and localized complexes in \TNS{} cause local distortions, which is in good agreement with the EI hypothesis in the former, but not in the latter. Further comparative studies of \TNS{} and \TNSe{}, in particular of their electronic properties, are required to confirm this hypothesis.

\section{Acknowledgments}
	We gratefully acknowledge A.~N.~Yaresko and S.~Kaiser for fruitful discussion. This work was partly supported by the Japan Society for the Promotion of Science (JSPS) KAKENHI (Grants No. 17H01140, JP15H05852, JP15K21717) and the Alexander von Humboldt foundation.


\begin{thebibliography}{18}%
\makeatletter
\providecommand \@ifxundefined [1]{%
 \@ifx{#1\undefined}
}%
\providecommand \@ifnum [1]{%
 \ifnum #1\expandafter \@firstoftwo
 \else \expandafter \@secondoftwo
 \fi
}%
\providecommand \@ifx [1]{%
 \ifx #1\expandafter \@firstoftwo
 \else \expandafter \@secondoftwo
 \fi
}%
\providecommand \natexlab [1]{#1}%
\providecommand \enquote  [1]{``#1''}%
\providecommand \bibnamefont  [1]{#1}%
\providecommand \bibfnamefont [1]{#1}%
\providecommand \citenamefont [1]{#1}%
\providecommand \href@noop [0]{\@secondoftwo}%
\providecommand \href [0]{\begingroup \@sanitize@url \@href}%
\providecommand \@href[1]{\@@startlink{#1}\@@href}%
\providecommand \@@href[1]{\endgroup#1\@@endlink}%
\providecommand \@sanitize@url [0]{\catcode `\\12\catcode `\$12\catcode
  `\&12\catcode `\#12\catcode `\^12\catcode `\_12\catcode `\%12\relax}%
\providecommand \@@startlink[1]{}%
\providecommand \@@endlink[0]{}%
\providecommand \url  [0]{\begingroup\@sanitize@url \@url }%
\providecommand \@url [1]{\endgroup\@href {#1}{\urlprefix }}%
\providecommand \urlprefix  [0]{URL }%
\providecommand \Eprint [0]{\href }%
\providecommand \doibase [0]{http://dx.doi.org/}%
\providecommand \selectlanguage [0]{\@gobble}%
\providecommand \bibinfo  [0]{\@secondoftwo}%
\providecommand \bibfield  [0]{\@secondoftwo}%
\providecommand \translation [1]{[#1]}%
\providecommand \BibitemOpen [0]{}%
\providecommand \bibitemStop [0]{}%
\providecommand \bibitemNoStop [0]{.\EOS\space}%
\providecommand \EOS [0]{\spacefactor3000\relax}%
\providecommand \BibitemShut  [1]{\csname bibitem#1\endcsname}%
\let\auto@bib@innerbib\@empty
%</preamble>
\bibitem [{\citenamefont {Wakisaka}\ \emph {et~al.}(2009)\citenamefont
  {Wakisaka}, \citenamefont {Sudayama}, \citenamefont {Takubo}, \citenamefont
  {Mizokawa}, \citenamefont {Arita}, \citenamefont {Namatame}, \citenamefont
  {Taniguchi}, \citenamefont {Katayama}, \citenamefont {Nohara},\ and\
  \citenamefont {Takagi}}]{Takagi_EI}%
  \BibitemOpen
  \bibfield  {author} {\bibinfo {author} {\bibfnamefont {Y.}~\bibnamefont
  {Wakisaka}}, \bibinfo {author} {\bibfnamefont {T.}~\bibnamefont {Sudayama}},
  \bibinfo {author} {\bibfnamefont {K.}~\bibnamefont {Takubo}}, \bibinfo
  {author} {\bibfnamefont {T.}~\bibnamefont {Mizokawa}}, \bibinfo {author}
  {\bibfnamefont {M.}~\bibnamefont {Arita}}, \bibinfo {author} {\bibfnamefont
  {H.}~\bibnamefont {Namatame}}, \bibinfo {author} {\bibfnamefont
  {M.}~\bibnamefont {Taniguchi}}, \bibinfo {author} {\bibfnamefont
  {N.}~\bibnamefont {Katayama}}, \bibinfo {author} {\bibfnamefont
  {M.}~\bibnamefont {Nohara}}, \ and\ \bibinfo {author} {\bibfnamefont
  {H.}~\bibnamefont {Takagi}},\ }\href@noop {} {\bibfield  {journal} {\bibinfo
  {journal} {Phys. Rev. Lett.}\ }\textbf {\bibinfo {volume} {103}},\ \bibinfo
  {pages} {026402} (\bibinfo {year} {2009})}\BibitemShut {NoStop}%
\bibitem [{\citenamefont {Wakisaka}\ \emph {et~al.}(2012)\citenamefont
  {Wakisaka}, \citenamefont {Sudayama}, \citenamefont {Takubo}, \citenamefont
  {Mizokawa}, \citenamefont {Saini}, \citenamefont {Arita}, \citenamefont
  {Namatame}, \citenamefont {Taniguchi}, \citenamefont {Katayama},
  \citenamefont {Nohara},\ and\ \citenamefont {Takagi}}]{Wakisaka2012}%
  \BibitemOpen
  \bibfield  {author} {\bibinfo {author} {\bibfnamefont {Y.}~\bibnamefont
  {Wakisaka}}, \bibinfo {author} {\bibfnamefont {T.}~\bibnamefont {Sudayama}},
  \bibinfo {author} {\bibfnamefont {K.}~\bibnamefont {Takubo}}, \bibinfo
  {author} {\bibfnamefont {T.}~\bibnamefont {Mizokawa}}, \bibinfo {author}
  {\bibfnamefont {N.~L.}\ \bibnamefont {Saini}}, \bibinfo {author}
  {\bibfnamefont {M.}~\bibnamefont {Arita}}, \bibinfo {author} {\bibfnamefont
  {H.}~\bibnamefont {Namatame}}, \bibinfo {author} {\bibfnamefont
  {M.}~\bibnamefont {Taniguchi}}, \bibinfo {author} {\bibfnamefont
  {N.}~\bibnamefont {Katayama}}, \bibinfo {author} {\bibfnamefont
  {M.}~\bibnamefont {Nohara}}, \ and\ \bibinfo {author} {\bibfnamefont
  {H.}~\bibnamefont {Takagi}},\ }\href {\doibase 10.1007/s10948-012-1526-0}
  {\bibfield  {journal} {\bibinfo  {journal} {J. Supercond.
  Nov. Magn.}\ }\textbf {\bibinfo {volume} {25}},\ \bibinfo {pages} {1231}
  (\bibinfo {year} {2012})}\BibitemShut {NoStop}%
\bibitem [{\citenamefont {Kaneko}\ \emph {et~al.}(2013)\citenamefont {Kaneko},
  \citenamefont {Toriyama}, \citenamefont {Konishi},\ and\ \citenamefont
  {Ohta}}]{KanekoTheory}%
  \BibitemOpen
  \bibfield  {author} {\bibinfo {author} {\bibfnamefont {T.}~\bibnamefont
  {Kaneko}}, \bibinfo {author} {\bibfnamefont {T.}~\bibnamefont {Toriyama}},
  \bibinfo {author} {\bibfnamefont {T.}~\bibnamefont {Konishi}}, \ and\
  \bibinfo {author} {\bibfnamefont {Y.}~\bibnamefont {Ohta}},\ }\href {\doibase
  10.1103/PhysRevB.87.035121} {\bibfield  {journal} {\bibinfo  {journal} {Phys.
  Rev. B}\ }\textbf {\bibinfo {volume} {87}},\ \bibinfo {pages} {035121}
  (\bibinfo {year} {2013})}\BibitemShut {NoStop}%
\bibitem [{\citenamefont {Seki}\ \emph {et~al.}(2014)\citenamefont {Seki},
  \citenamefont {Wakisaka}, \citenamefont {Kaneko}, \citenamefont {Toriyama},
  \citenamefont {Konishi}, \citenamefont {Sudayama}, \citenamefont {Saini},
  \citenamefont {Arita}, \citenamefont {Namatame}, \citenamefont {Taniguchi},
  \citenamefont {Katayama}, \citenamefont {Nohara}, \citenamefont {Takagi},
  \citenamefont {Mizokawa},\ and\ \citenamefont {Ohta}}]{BEC_in_TNSe}%
  \BibitemOpen
  \bibfield  {author} {\bibinfo {author} {\bibfnamefont {K.}~\bibnamefont
  {Seki}}, \bibinfo {author} {\bibfnamefont {Y.}~\bibnamefont {Wakisaka}},
  \bibinfo {author} {\bibfnamefont {T.}~\bibnamefont {Kaneko}}, \bibinfo
  {author} {\bibfnamefont {T.}~\bibnamefont {Toriyama}}, \bibinfo {author}
  {\bibfnamefont {T.}~\bibnamefont {Konishi}}, \bibinfo {author} {\bibfnamefont
  {T.}~\bibnamefont {Sudayama}}, \bibinfo {author} {\bibfnamefont {N.~L.}\
  \bibnamefont {Saini}}, \bibinfo {author} {\bibfnamefont {M.}~\bibnamefont
  {Arita}}, \bibinfo {author} {\bibfnamefont {H.}~\bibnamefont {Namatame}},
  \bibinfo {author} {\bibfnamefont {M.}~\bibnamefont {Taniguchi}}, \bibinfo
  {author} {\bibfnamefont {N.}~\bibnamefont {Katayama}}, \bibinfo {author}
  {\bibfnamefont {M.}~\bibnamefont {Nohara}}, \bibinfo {author} {\bibfnamefont
  {H.}~\bibnamefont {Takagi}}, \bibinfo {author} {\bibfnamefont
  {T.}~\bibnamefont {Mizokawa}}, \ and\ \bibinfo {author} {\bibfnamefont
  {Y.}~\bibnamefont {Ohta}},\ }\href {\doibase 10.1103/PhysRevB.90.155116}
  {\bibfield  {journal} {\bibinfo  {journal} {Phys. Rev. B}\ }\textbf {\bibinfo
  {volume} {90}},\ \bibinfo {pages} {155116} (\bibinfo {year}
  {2014})}\BibitemShut {NoStop}%
\bibitem [{\citenamefont {Lu}\ \emph {et~al.}(2017)\citenamefont {Lu},
  \citenamefont {Kono}, \citenamefont {Larkin}, \citenamefont {Rost},
  \citenamefont {Takayama}, \citenamefont {Boris}, \citenamefont {Keimer},\
  and\ \citenamefont {Takagi}}]{Takagi2016}%
  \BibitemOpen
  \bibfield  {author} {\bibinfo {author} {\bibfnamefont {Y.-F.}\ \bibnamefont
  {Lu}}, \bibinfo {author} {\bibfnamefont {H.}~\bibnamefont {Kono}}, \bibinfo
  {author} {\bibfnamefont {T.~I.}\ \bibnamefont {Larkin}}, \bibinfo {author}
  {\bibfnamefont {A.~W.}\ \bibnamefont {Rost}}, \bibinfo {author}
  {\bibfnamefont {T.}~\bibnamefont {Takayama}}, \bibinfo {author}
  {\bibfnamefont {A.~V.}\ \bibnamefont {Boris}}, \bibinfo {author}
  {\bibfnamefont {B.}~\bibnamefont {Keimer}}, \ and\ \bibinfo {author}
  {\bibfnamefont {H.}~\bibnamefont {Takagi}},\ }\href@noop {} {\bibfield
  {journal} {\bibinfo  {journal} {Nature Communications}\ }\textbf {\bibinfo
  {volume} {8}},\ \bibinfo {pages} {14408} (\bibinfo {year}
  {2017})}\BibitemShut {NoStop}%
\bibitem [{\citenamefont {Larkin}\ \emph {et~al.}(2017)\citenamefont {Larkin},
  \citenamefont {Yaresko}, \citenamefont {Pr{\"o}pper}, \citenamefont {Kikoin},
  \citenamefont {Lu}, \citenamefont {Takayama}, \citenamefont {Mathis},
  \citenamefont {Rost}, \citenamefont {Takagi}, \citenamefont {Keimer},\ and\
  \citenamefont {Boris}}]{Larkin2017}%
  \BibitemOpen
  \bibfield  {author} {\bibinfo {author} {\bibfnamefont {T.~I.}\ \bibnamefont
  {Larkin}}, \bibinfo {author} {\bibfnamefont {A.~N.}\ \bibnamefont {Yaresko}},
  \bibinfo {author} {\bibfnamefont {D.}~\bibnamefont {Pr{\"o}pper}}, \bibinfo
  {author} {\bibfnamefont {K.~A.}\ \bibnamefont {Kikoin}}, \bibinfo {author}
  {\bibfnamefont {Y.-F.}\ \bibnamefont {Lu}}, \bibinfo {author} {\bibfnamefont
  {T.}~\bibnamefont {Takayama}}, \bibinfo {author} {\bibfnamefont {Y.-L.}\
  \bibnamefont {Mathis}}, \bibinfo {author} {\bibfnamefont {A.~W.}\
  \bibnamefont {Rost}}, \bibinfo {author} {\bibfnamefont {H.}~\bibnamefont
  {Takagi}}, \bibinfo {author} {\bibfnamefont {B.}~\bibnamefont {Keimer}}, \
  and\ \bibinfo {author} {\bibfnamefont {A.~V.}\ \bibnamefont {Boris}},\
  }\href@noop {} {\bibfield  {journal} {\bibinfo  {journal} {Phys. Rev. B}\
  }\textbf {\bibinfo {volume} {95}},\ \bibinfo {pages} {195144} (\bibinfo
  {year} {2017})}\BibitemShut {NoStop}%
\bibitem [{\citenamefont {Werdehausen}\ \emph {et~al.}(2018)\citenamefont
  {Werdehausen}, \citenamefont {Takayama}, \citenamefont {H{\"o}ppner},
  \citenamefont {Albrecht}, \citenamefont {Rost}, \citenamefont {Lu},
  \citenamefont {Manske}, \citenamefont {Takagi},\ and\ \citenamefont
  {Kaiser}}]{Werdehausen2018}%
  \BibitemOpen
  \bibfield  {author} {\bibinfo {author} {\bibfnamefont {D.}~\bibnamefont
  {Werdehausen}}, \bibinfo {author} {\bibfnamefont {T.}~\bibnamefont
  {Takayama}}, \bibinfo {author} {\bibfnamefont {M.}~\bibnamefont
  {H{\"o}ppner}}, \bibinfo {author} {\bibfnamefont {G.}~\bibnamefont
  {Albrecht}}, \bibinfo {author} {\bibfnamefont {A.~W.}\ \bibnamefont {Rost}},
  \bibinfo {author} {\bibfnamefont {Y.}~\bibnamefont {Lu}}, \bibinfo {author}
  {\bibfnamefont {D.}~\bibnamefont {Manske}}, \bibinfo {author} {\bibfnamefont
  {H.}~\bibnamefont {Takagi}}, \ and\ \bibinfo {author} {\bibfnamefont
  {S.}~\bibnamefont {Kaiser}},\ }\href@noop {} {\bibfield  {journal} {\bibinfo
  {journal} {Science Advances}\ }\textbf {\bibinfo {volume} {4}} (\bibinfo
  {year} {2018})}\BibitemShut {NoStop}%
\bibitem [{\citenamefont {Bucher}\ \emph {et~al.}(1991)\citenamefont {Bucher},
  \citenamefont {Steiner},\ and\ \citenamefont {Wachter}}]{TmSeTe_Pressure_91}%
  \BibitemOpen
  \bibfield  {author} {\bibinfo {author} {\bibfnamefont {B.}~\bibnamefont
  {Bucher}}, \bibinfo {author} {\bibfnamefont {P.}~\bibnamefont {Steiner}}, \
  and\ \bibinfo {author} {\bibfnamefont {P.}~\bibnamefont {Wachter}},\ }\href
  {\doibase 10.1103/PhysRevLett.67.2717} {\bibfield  {journal} {\bibinfo
  {journal} {Phys. Rev. Lett.}\ }\textbf {\bibinfo {volume} {67}},\ \bibinfo
  {pages} {2717} (\bibinfo {year} {1991})}\BibitemShut {NoStop}%
\bibitem [{\citenamefont {Wachter}\ and\ \citenamefont
  {Bucher}(2013)}]{WachterBands}%
  \BibitemOpen
  \bibfield  {author} {\bibinfo {author} {\bibfnamefont {P.}~\bibnamefont
  {Wachter}}\ and\ \bibinfo {author} {\bibfnamefont {B.}~\bibnamefont
  {Bucher}},\ }\href {\doibase http://dx.doi.org/10.1016/j.physb.2012.09.018}
  {\bibfield  {journal} {\bibinfo  {journal} {Physica B: Condensed Matter}\
  }\textbf {\bibinfo {volume} {408}},\ \bibinfo {pages} {51 } (\bibinfo {year}
  {2013})}\BibitemShut {NoStop}%
\bibitem [{\citenamefont {Monney}\ \emph {et~al.}(2011)\citenamefont {Monney},
  \citenamefont {Battaglia}, \citenamefont {Cercellier}, \citenamefont {Aebi},\
  and\ \citenamefont {Beck}}]{TiSe2_ARPES5}%
  \BibitemOpen
  \bibfield  {author} {\bibinfo {author} {\bibfnamefont {C.}~\bibnamefont
  {Monney}}, \bibinfo {author} {\bibfnamefont {C.}~\bibnamefont {Battaglia}},
  \bibinfo {author} {\bibfnamefont {H.}~\bibnamefont {Cercellier}}, \bibinfo
  {author} {\bibfnamefont {P.}~\bibnamefont {Aebi}}, \ and\ \bibinfo {author}
  {\bibfnamefont {H.}~\bibnamefont {Beck}},\ }\href {\doibase
  10.1103/PhysRevLett.106.106404} {\bibfield  {journal} {\bibinfo  {journal}
  {Phys. Rev. Lett.}\ }\textbf {\bibinfo {volume} {106}},\ \bibinfo {pages}
  {106404} (\bibinfo {year} {2011})}\BibitemShut {NoStop}%
\bibitem [{\citenamefont {Sunshine}\ and\ \citenamefont
  {Ibers}(1985)}]{SunshineIbers1985}%
  \BibitemOpen
  \bibfield  {author} {\bibinfo {author} {\bibfnamefont {S.~A.}\ \bibnamefont
  {Sunshine}}\ and\ \bibinfo {author} {\bibfnamefont {J.~A.}\ \bibnamefont
  {Ibers}},\ }\href {\doibase 10.1021/ic00216a027} {\bibfield  {journal}
  {\bibinfo  {journal} {Inorg. Chem.}\ }\textbf {\bibinfo {volume} {24}},\
  \bibinfo {pages} {3611} (\bibinfo {year} {1985})}\BibitemShut {NoStop}%
\bibitem [{\citenamefont {Baroni}\ \emph {et~al.}(2001)\citenamefont {Baroni},
  \citenamefont {de~Gironcoli}, \citenamefont {Dal~Corso},\ and\ \citenamefont
  {Giannozzi}}]{baroni_dfpt}%
  \BibitemOpen
  \bibfield  {author} {\bibinfo {author} {\bibfnamefont {S.}~\bibnamefont
  {Baroni}}, \bibinfo {author} {\bibfnamefont {S.}~\bibnamefont
  {de~Gironcoli}}, \bibinfo {author} {\bibfnamefont {A.}~\bibnamefont
  {Dal~Corso}}, \ and\ \bibinfo {author} {\bibfnamefont {P.}~\bibnamefont
  {Giannozzi}},\ }\href {\doibase 10.1103/RevModPhys.73.515} {\bibfield
  {journal} {\bibinfo  {journal} {Rev. Mod. Phys.}\ }\textbf {\bibinfo {volume}
  {73}},\ \bibinfo {pages} {515} (\bibinfo {year} {2001})}\BibitemShut
  {NoStop}%
\bibitem [{\citenamefont {Giannozzi}\ \emph {et~al.}(2009)\citenamefont
  {Giannozzi}, \citenamefont {Baroni}, \citenamefont {Bonini}, \citenamefont
  {Calandra}, \citenamefont {Car}, \citenamefont {Cavazzoni}, \citenamefont
  {Ceresoli}, \citenamefont {Chiarotti}, \citenamefont {Cococcioni},
  \citenamefont {Dabo}, \citenamefont {Corso}, \citenamefont {de~Gironcoli},
  \citenamefont {Fabris}, \citenamefont {Fratesi}, \citenamefont {Gebauer},
  \citenamefont {Gerstmann}, \citenamefont {Gougoussis}, \citenamefont
  {Kokalj}, \citenamefont {Lazzeri}, \citenamefont {Martin-Samos},
  \citenamefont {Marzari}, \citenamefont {Mauri}, \citenamefont {Mazzarello},
  \citenamefont {Paolini}, \citenamefont {Pasquarello}, \citenamefont
  {Paulatto}, \citenamefont {Sbraccia}, \citenamefont {Scandolo}, \citenamefont
  {Sclauzero}, \citenamefont {Seitsonen}, \citenamefont {Smogunov},
  \citenamefont {Umari},\ and\ \citenamefont {Wentzcovitch}}]{qe}%
  \BibitemOpen
  \bibfield  {author} {\bibinfo {author} {\bibfnamefont {P.}~\bibnamefont
  {Giannozzi}}, \bibinfo {author} {\bibfnamefont {S.}~\bibnamefont {Baroni}},
  \bibinfo {author} {\bibfnamefont {N.}~\bibnamefont {Bonini}}, \bibinfo
  {author} {\bibfnamefont {M.}~\bibnamefont {Calandra}}, \bibinfo {author}
  {\bibfnamefont {R.}~\bibnamefont {Car}}, \bibinfo {author} {\bibfnamefont
  {C.}~\bibnamefont {Cavazzoni}}, \bibinfo {author} {\bibfnamefont
  {D.}~\bibnamefont {Ceresoli}}, \bibinfo {author} {\bibfnamefont {G.~L.}\
  \bibnamefont {Chiarotti}}, \bibinfo {author} {\bibfnamefont {M.}~\bibnamefont
  {Cococcioni}}, \bibinfo {author} {\bibfnamefont {I.}~\bibnamefont {Dabo}},
  \bibinfo {author} {\bibfnamefont {A.~D.}\ \bibnamefont {Corso}}, \bibinfo
  {author} {\bibfnamefont {S.}~\bibnamefont {de~Gironcoli}}, \bibinfo {author}
  {\bibfnamefont {S.}~\bibnamefont {Fabris}}, \bibinfo {author} {\bibfnamefont
  {G.}~\bibnamefont {Fratesi}}, \bibinfo {author} {\bibfnamefont
  {R.}~\bibnamefont {Gebauer}}, \bibinfo {author} {\bibfnamefont
  {U.}~\bibnamefont {Gerstmann}}, \bibinfo {author} {\bibfnamefont
  {C.}~\bibnamefont {Gougoussis}}, \bibinfo {author} {\bibfnamefont
  {A.}~\bibnamefont {Kokalj}}, \bibinfo {author} {\bibfnamefont
  {M.}~\bibnamefont {Lazzeri}}, \bibinfo {author} {\bibfnamefont
  {L.}~\bibnamefont {Martin-Samos}}, \bibinfo {author} {\bibfnamefont
  {N.}~\bibnamefont {Marzari}}, \bibinfo {author} {\bibfnamefont
  {F.}~\bibnamefont {Mauri}}, \bibinfo {author} {\bibfnamefont
  {R.}~\bibnamefont {Mazzarello}}, \bibinfo {author} {\bibfnamefont
  {S.}~\bibnamefont {Paolini}}, \bibinfo {author} {\bibfnamefont
  {A.}~\bibnamefont {Pasquarello}}, \bibinfo {author} {\bibfnamefont
  {L.}~\bibnamefont {Paulatto}}, \bibinfo {author} {\bibfnamefont
  {C.}~\bibnamefont {Sbraccia}}, \bibinfo {author} {\bibfnamefont
  {S.}~\bibnamefont {Scandolo}}, \bibinfo {author} {\bibfnamefont
  {G.}~\bibnamefont {Sclauzero}}, \bibinfo {author} {\bibfnamefont {A.~P.}\
  \bibnamefont {Seitsonen}}, \bibinfo {author} {\bibfnamefont {A.}~\bibnamefont
  {Smogunov}}, \bibinfo {author} {\bibfnamefont {P.}~\bibnamefont {Umari}}, \
  and\ \bibinfo {author} {\bibfnamefont {R.~M.}\ \bibnamefont {Wentzcovitch}},\
  }\href {http://stacks.iop.org/0953-8984/21/i=39/a=395502} {\bibfield
  {journal} {\bibinfo  {journal} {Journal of Physics: Condensed Matter}\
  }\textbf {\bibinfo {volume} {21}},\ \bibinfo {pages} {395502} (\bibinfo
  {year} {2009})}\BibitemShut {NoStop}%
\bibitem [{\citenamefont {Bl\"ochl}(1994)}]{paw_bloechl}%
  \BibitemOpen
  \bibfield  {author} {\bibinfo {author} {\bibfnamefont {P.~E.}\ \bibnamefont
  {Bl\"ochl}},\ }\href {\doibase 10.1103/PhysRevB.50.17953} {\bibfield
  {journal} {\bibinfo  {journal} {Phys. Rev. B}\ }\textbf {\bibinfo {volume}
  {50}},\ \bibinfo {pages} {17953} (\bibinfo {year} {1994})}\BibitemShut
  {NoStop}%
\bibitem [{\citenamefont {Corso}(2014)}]{pslibrary}%
  \BibitemOpen
  \bibfield  {author} {\bibinfo {author} {\bibfnamefont {A.~D.}\ \bibnamefont
  {Corso}},\ }\href {\doibase https://doi.org/10.1016/j.commatsci.2014.07.043}
  {\bibfield  {journal} {\bibinfo  {journal} {Computational Materials Science}\
  }\textbf {\bibinfo {volume} {95}},\ \bibinfo {pages} {337 } (\bibinfo {year}
  {2014})}\BibitemShut {NoStop}%
\bibitem [{pse()}]{pseudos}%
  \BibitemOpen
  \href@noop {} {}\bibinfo {note} {We used the paw pseudo potentials from
  pslibrary 0.3.1}\BibitemShut {NoStop}%
\bibitem [{\citenamefont {Schubert}\ \emph {et~al.}(2000)\citenamefont
  {Schubert}, \citenamefont {Tiwald},\ and\ \citenamefont
  {Herzinger}}]{Schubert2000}%
  \BibitemOpen
  \bibfield  {author} {\bibinfo {author} {\bibfnamefont {M.}~\bibnamefont
  {Schubert}}, \bibinfo {author} {\bibfnamefont {T.~E.}\ \bibnamefont
  {Tiwald}}, \ and\ \bibinfo {author} {\bibfnamefont {C.~M.}\ \bibnamefont
  {Herzinger}},\ }\href {\doibase 10.1103/PhysRevB.61.8187} {\bibfield
  {journal} {\bibinfo  {journal} {Phys. Rev. B}\ }\textbf {\bibinfo {volume}
  {61}},\ \bibinfo {pages} {8187} (\bibinfo {year} {2000})}\BibitemShut
  {NoStop}%
\bibitem [{\citenamefont {Huml\'{\i}\ifmmode~\check{c}\else \v{c}\fi{}ek}\
  \emph {et~al.}(2000)\citenamefont {Huml\'{\i}\ifmmode~\check{c}\else
  \v{c}\fi{}ek}, \citenamefont {Henn},\ and\ \citenamefont
  {Cardona}}]{Humlicek2000}%
  \BibitemOpen
  \bibfield  {author} {\bibinfo {author} {\bibfnamefont {J.}~\bibnamefont
  {Huml\'{\i}\ifmmode~\check{c}\else \v{c}\fi{}ek}}, \bibinfo {author}
  {\bibfnamefont {R.}~\bibnamefont {Henn}}, \ and\ \bibinfo {author}
  {\bibfnamefont {M.}~\bibnamefont {Cardona}},\ }\href {\doibase
  10.1103/PhysRevB.61.14554} {\bibfield  {journal} {\bibinfo  {journal} {Phys.
  Rev. B}\ }\textbf {\bibinfo {volume} {61}},\ \bibinfo {pages} {14554}
  (\bibinfo {year} {2000})}\BibitemShut {NoStop}%
\end{thebibliography}
\end{document}